\documentclass[twocolumn,prl,showpacs,preprintnumbers,amsmath,amssymb]{revtex4}
\usepackage{bm}
\input{epsf}

\begin{document}
\title{Properties of Bose gas in a lattice model \\ (strong
interaction)}
\author{E.G. Batyev}\email{batyev@isp.nsc.ru}\affiliation{A.V. Rzhanov Institute of Semiconductor
Physics , 630090 Novosibirsk, Russia}

\begin{abstract} The doubts concerning validity of gas approximation for
strong interaction (for example, hard spheres) are expressed. A
contradictory example - a Bose system in a lattice model - is
considered. Namely, the $X-Y$ model for spin $1/2$ is taken. A state
with spins directed downwards is considered to be vacuum with
respect to the particles. An inverse spin (+1/2) corresponds to a
particle. There is an usual band spectrum (one band) for a single
particle. A trial function is written for a multiple particle
system. This function was shown to be a good one within a
macroscopic limit, when the particle and lattice site numbers tend
to infinity (the ratio of energy standard deviation to average
energy tends to zero). A result within the gas limit (particle
number is small compared with the lattice site number) is compared
with that obtained via generally accepted approach.
\end{abstract}
\maketitle
\section{Introduction}
As is known, a Bose gas model with a weak interaction (repulsion),
considered by Bogoliubov \cite{1}, is generalized for a case of an
arbitrary interaction value provided that a so-called gas
approximation is correct (for example, see \cite{2}). The main point
of  this generalization is a transition from true interaction to
scattering length (which is small in comparison with inter-particle
distance within the gas approximation). However, the gas
approximation, being good for a classical case, is not always
suitable for a quantum one. The reason is the absence of a
conception about trajectories (no quasi-classics) in ultra-quantum
limit (which takes place for Bose system). Therefore one can hardly
speak about binary collisions. In fact, interaction of every
particle with all the particles at once is rather probable: when
particle number is small, most particles are in Bose condensate,
i.e. these particles are characterized by an infinite wave length.
The aim of the present work is to show this phenomenon within the
framework of a simple model. Namely, a Bose gas in the lattice model
with infinite interaction is considered, so that no more than one
particle can be at each cite (description of interaction like for
hard spheres). This model is equivalent to X-Y model for spin $1/2$
, where particle vacuum is a state with all spins directed downwards
and spin directed upwards $(+1/2)$ corresponds to a particle. An
average energy value is calculated using a trial function at given
particle number written for the X-Y model and a ratio of a
mean-square energy deviation to an average energy value is shown to
tend to zero within a macroscopic limit. The result of calculation
of the basic state energy via the traditional approach (according to
the accepted rules) is shown to be different.

Bose gas in the hard spheres model was considered, for example, in
the work [3], where pseudo-potential method was used. The
pseudo-potential value is selected so that the scattering of two
particles at each other was the same as in the case of hard spheres.
It is the pseudo-potential, for which the corresponding
multi-particle Hamiltonian is written, then the interaction is
presented in a more simple form and, finally, the basic state energy
value (calculated from this simple form) is given ([3], section 1).
This model and mathematical treatment is given in [4] also.
Nevertheless, no accuracy estimate of the traditional approach was
made as nobody used the trial function for initial interaction (hard
spheres). This raises the question of whether this approach is
correct in the case of Bose condensate, when no conception about
particle trajectory and accordingly about binary collisions exists.

\section{The model}

The Hamiltonian of the $X-Y$ model is: \begin{eqnarray} \label{1}H =
 -t\sum_{<nn'>}S^+(n)S^-({n'})\ \ \ \ \ (t>0)\ .\end{eqnarray}
Here $<nn'>$ denotes nearest neighbors (the sum takes place over
nearest neighbors), and operators $S^{\pm}(n)$ relate to spin $1/2$
at the site with number $n$:
$$S^{\pm}(n) = S_x(n)\pm iS_y(n)$$ (corresponding radius - vector of the site is ${\bf R}_n$).
These operators commute at different sites and at one site we have:
\begin{eqnarray} \label{2}S^-(n)S^+(n)-S^+(n)S^-(n)=-2S_z(n)\ ,\\
\nonumber S_z(n)S^+(n)-S^+(n)S_z(n)=S^+(n)\ . \end{eqnarray}

The following Hamiltonian can be written for the Bose particles in
the lattice model instead of  (\ref{1}):  \begin{eqnarray}\label{3}
H\rightarrow -t\sum_{<nn'>}A_n^+A_{n'} +\\ \nonumber
+U\sum_nA_n^+A_n^+A_nA_n \ \ \ \ \ \  (U\rightarrow\infty)\
.\end{eqnarray}

Operators $A_n^+,\ A_n$ are the operators of creation and
destruction of a Bose particle at a site with number n. It is
obvious that these statements of the problem are equivalent. Further
the spin approach (\ref{1}) will be used mainly.

The state $\Phi_0$ with all spins directed down (-1/2) is taken as
initial one, i.e.: $$\Phi_0\equiv |\downarrow>\ \ \ \ \ (H\Phi_0=0)\
.$$ It is vacuum with respect to the particles. A wave function
$S^+(n)|\downarrow>$ corresponding to one inverse spin is equivalent
to a particle (a particle at a site with number n). As the state
$\Bigl(S^+(n)\Bigr)^2|\downarrow> =0$, there can be only one
particle at a site (two or more particles can not exist at one
site), i.e. it can be considered to be the model with interaction of
the hard spheres (equivalent of the condition $U\rightarrow\infty$
in expression (\ref{3})).

The spectrum $\epsilon$ of a single particle is found by a
conventional method: the wave function $\Phi_1$ has the form:
$$\Phi_1=\sum_nC_nS^+(n)|\downarrow>\ ,\ \ \ \ \ \ H\Phi_1=\epsilon\Phi_1\ .$$
Hence we have the following:
$$-t\sum_{<nm>}C_mS^+(n)|\downarrow> =
\epsilon\sum_nC_nS^+(n)|\downarrow>\ ,$$$$ \epsilon C_n
=-t\sum_{\nu} C_{n+\nu} \ .$$ The sum with respect to $\nu$ in the
last expression is the sum with respect to the nearest neighbors to
the site with number $n$. The solution in the form
$C_n\sim\exp\Bigl(i{\bf kR}_n\Bigr)$ is sought. Hence we have the
following: \begin{eqnarray} \label{4}\epsilon({\bf
k})=-t\sum_\nu\exp\Bigl(i{\bf k R}_n-i{\bf k
R}_{n+\nu}\Bigr)\rightarrow \\ \nonumber
-2t\Bigl\{\cos(k_xa)+\cos(k_ya)+\cos(k_za) \Bigr\}\ . \end{eqnarray}
Here the expression for a simple cubic lattice is presented ($a$ is
the lattice period). In the vicinity to the bottom of the band we
have:  $$\epsilon({\bf k})\approx -t\nu_0 +\frac{k^2}{2m}\ \ \ \ \ \
\ \Bigl(1/m=2ta^2\Bigr)\ .$$

The following values correspond to the ground
state of one particle:
$$\Phi_1\sim \sum_{n=1}^{N_0}S^+(n) |\downarrow>\ ,\ \ \ \
\epsilon(0)=-\nu_0 t $$ ($\nu_0$ is the number of nearest neighbors,
$N_0$ is the number of the lattice sites).

The value $\epsilon(0)$ is the beginning of the particle energy
counting (insignificant value). If the particle number $N$ is small
($N<<N_0$), one can talk about the Bose gas. The particle number
operator is \begin{eqnarray} \label{5}\hat{ N}=\frac{1}{2}\
\sum_{n=1}^{N_0} \Bigl(1+2S_z(n)\Bigr) = \frac{N_0}{2} + S_z \\
\nonumber \Bigl(S_z\equiv \sum_{n=1}^{N_0}S_z(n)\Bigr)\
.\end{eqnarray} It is integral of motion  (commutates with the
Hamiltonian). This is obvious from the Hamiltonian form and can be
directly confirmed.

\section{Trial function}

Above mentioned taken into account, the trial function has the
following form: \begin{eqnarray} \label{6} \Phi_N= \Bigl(
S^+\Bigr)^N |\downarrow> \ \ \ \ \ \ \ \ \ \ \biggl( S^+\equiv
\sum_{n=1}^{N_0}S^+(n)\biggr) \ . \end{eqnarray} This approximation
can be expected to be good (at least for small particle number
$N<<N_0$), as all the particles are in Bose condensate.

First let us consider normalization, i.e. the value
$(\Phi_N,\Phi_N)$. The initial state $|\downarrow>$ (vacuum to
particles) corresponds to maximum system spin ($N_0/2$) with maximum
negative projection ($-N_0/2$). This state is normalized. $S^+$
operator raises the projection by a unity without changing of the
full spin value. Matrix elements of the operator (calculated by the
normalized functions) are known from the general courses of quantum
mechanics: $$\Bigl( S^+\Bigr)_{M,M-1} = \sqrt{(S+M)(S-M+1)}\ .$$
Here $M$ is $S_z$ spin projection value. For example, the effect of
operation on the particle's vacuum ($M-1=-N_0/2,\ S=N_0/2$) is:
$$\Bigl( S^+\Bigr)_{M,M-1} = \sqrt{N_0}\ \ \rightarrow\ \ S^+\Phi_0
= \sqrt{N_0}\Phi_1 $$$$ (M=-N_0/2+1)\ .$$ Consequently, having
denoted the corresponding to $\Phi_N$ normalized function by
$\widetilde{\Phi}_N\ (\widetilde{\Phi}_N\equiv D_N\Phi_N)$, one can
derive: $$\frac{1}{D_N} = \prod_{M=1-N_0/2}^{N-N_0/2}
\sqrt{(S+M)(S-M+1)} =$$$$=\prod_{n=1}^N\sqrt{n(N_0-n+1)}\ . $$ And,
finally:  \begin{eqnarray} \label{7} \widetilde{\Phi}_N=D_N\Phi_N\
;\\ \nonumber D_N = \sqrt{\frac{(N_0-N)!}{N!N_0!}}\ .\end{eqnarray}

\subsection{Auxiliary relations} Some relations are necessary in what
follows. The most simple is the calculation of, for example,
$\Bigl(\Phi_N,\Phi_N(n) \Bigr)$, where $\Phi_N(n)\equiv
S^+(n)\Phi_{N-1}$. This value does not depend on the cite number
$n$, as all the cites are equivalent. Therefore one can write:
$$\Phi_N(n)\equiv S^+(n)\Phi_{N-1}\ ;$$
\begin{eqnarray} \label{8} \Bigl(\Phi_N,\Phi_N(n) \Bigr) =
\frac{1}{N_0}\sum_n \Bigl(\Phi_N,\Phi_N(n) \Bigr) =\frac{1}{N_0} (\Phi_N,\Phi_N) ,\\
\nonumber \Bigl(\Phi_N,\Phi_N(n,n') \Bigr) = \frac{1}{N_0(N_0-1)}
(\Phi_N,\Phi_N)\ .\end{eqnarray} Analogous relation can be written
for the case of both functions containing the cite number, for
example:
$$\Bigl(\Phi_N(n'),\Phi_N(n) \Bigr)_{n\neq n'} =
\frac{1}{N_0-1}\biggl\{\sum_{n'}\Bigl(\Phi_N(n'),\Phi_N(n)
\Bigr)-$$$$ - \Bigl(\Phi_N(n),\Phi_N(n) \Bigr)\biggr\} = $$ $$=\
\frac{1}{N_0-1}\biggl\{\Bigl(\Phi_N,\Phi_N(n) \Bigr)
-\Bigl(\Phi_N(n),\Phi_N(n) \Bigr) \biggr\}\ .$$ As for the value
with coinciding numbers ($n'=n$), one can notice that one of these
states is occupied a fortiori, so a norm of the state is got using
the relation (\ref{7}) and substituting $N_0\rightarrow (N_0-1),\ N
\rightarrow (N-1)$. The result is the following:  \begin{eqnarray}
\label{9} \Bigl(\Phi_N(n),\Phi_N(n) \Bigr) =
\frac{1}{NN_0}(\Phi_N,\Phi_N)\ ;\\ \nonumber
\Bigl(\Phi_N(n'),\Phi_N(n) \Bigr)_{n\neq n'} =
\frac{N-1}{NN_0(N_0-1)}\Bigl(\Phi_N,\Phi_N \Bigr)\ .\end{eqnarray}
The correctness of the relation is tested by substitution of $N=1$.

\section{ Energy}
Now an average energy value can be calculated: $$E =
(\widetilde{\Phi}_N,H\widetilde{\Phi}_N) = D_N^2(\Phi_N,H\Phi_N) =
$$$$=-tD_N^2N_0\nu_0\ \Bigl(\Phi_{N+1}(n'),\Phi_{N+1}(n) \Bigr)_{n'\neq
n}\ .$$ Hence taking into account (\ref{7}), (\ref{9}) the following
relation is derived: \begin{eqnarray} \label{10} E = -tD_N^2\
\frac{N \nu_0}{(N+1)(N_0-1)}\ \Bigl(\Phi_{N+1},\Phi_{N+1} \Bigr) =
\\ \nonumber =(-t\nu_0)\ N\Biggl\{1-\frac{N-1}{N_0-1} \Biggr\}\ .\end{eqnarray}
Note natural symmetry at the substitution $N\rightarrow(N_0-N)$. The
contribution linear in relation to the particle number $N$ is just
particle energy at the band bottom, quadratic contribution is a
result of particle interaction being taken into account (the test of
correctness: true result after substitution of $N$ by $1$).

It is easy to see, that similar energy value is obtained by
simplified approach, namely, when using the trial function in the
form: \begin{eqnarray} \label{11}\Phi^{(0)} = \prod_{n=1}^{N_0}
(u+vS^+(n))|\downarrow>\\ \nonumber (u^2+v^2 = 1\ , \ \ v^2=N/N_0)\
;\end{eqnarray} $$E^{(0)}=(\Phi^{(0)},H\Phi^{(0)})=
(-tN_0\nu_0)(uv)^2 =$$$$=(-t\nu_0)\ N\Biggl\{1-\frac{N}{N_0}
\Biggr\}\ .$$ The function (\ref{11}) is a self-consistent field
approximation.

It is interesting to note, that foregoing is true in a
two-dimensional case (square lattice and three-dimensional spin).

\subsection{Distribution function}

A distribution function can be found for the state (\ref{11}). The
trial function (\ref{11}) can be rewritten using Bose particles and
presented in the form: $$\Phi^{(0)} \rightarrow \prod_{n=1}^{N_0}
(u+vA^+_n)|0>\ .$$ The number of particles with given quasi-momentum
is: $$<A^+({\bf k})A({\bf k})> =
\frac{1}{N_0}\sum_{n,n'}<A^+_nA_{n'}>\times$$$$\times \exp\biggl\{
i{\bf k}\Bigl[{\bf R}(n')- {\bf R}(n)\Bigr] \biggr\} ;$$$$
\sum_{n,n'} =\sum_{n\neq n'}+\sum_{n=n'}\ . $$

The forbidding of two(many)fold occupation of the cites should be
taken into account. The result of the calculation is:
\begin{eqnarray} \label{12} n(0)=<A^+(0)A(0)> = N\biggl(1-
\frac{N}{N_0}\biggr)+ \biggl( \frac{N}{N_0}  \biggr)^2\ ,\\
\nonumber n({\bf k})= <A^+({\bf k})A({\bf k)}>\Bigl|_{{\bf k}\neq
0}=\biggl( \frac{N}{N_0}  \biggr)^2\ .\end{eqnarray} The summation
gives the required result:  $$ <A^+(0)A(0)>+\sum_{{\bf k}\neq
0}<A^+({\bf k})A({\bf k})> = N$$ (state number $N_0-1$ should be
taken into account in the sum over ${\bf k}\neq 0$). Then the energy
value is the same: $$E^{(0)} \rightarrow -t\nu_0 <A^+(0)A(0)> +
\sum_{{\bf k}\neq 0}\epsilon({\bf k})<A^+({\bf k})A({\bf k})> .$$

It should be emphasized, that though most particles are in the
condensate (see (\ref{12})), the approximate wave function of the
system cannot be written in the form $(A^+(0))^N|0>$ (in contrast to
weak interaction \cite{1}). Otherwise, the forbidden case can take
place, i.e. the particles can meet at one cite.

\subsection{The traditional approach}

    It is interesting to compare obtained energy value with the one obtained
using the traditional approach (see \cite{2}) within the gas
approximation (in our case it takes place at  $N<<N_0$). The system
energy can be estimated within the gas approximation using the
scattering amplitude. This means to find a vertex function in stair
approximation and then to write interaction energy in main
approximation, provided that all the particles are in the
condensate.

For this purpose Bose particles and their interaction according to
Hubbard is used (see (\ref{3})). The relation for interaction energy
is:  \begin{eqnarray} \label{13} H_{int} = U\sum_n A^+_n A^+_nA_nA_n
=\ \ \ \ \ \ \ \ \ \\ \nonumber =\frac{U}{N_0}\sum_{{\bf p}_1+{\bf
p}_2={\bf p}_3+{\bf p}_4} A^+({\bf p}_1)A^+({\bf p}_2)A({\bf
p}_3)A({\bf p}_4)\Bigl|_{U\rightarrow\infty} .\end{eqnarray}

According to \cite{2} a full vertex function $\Gamma$, describing
mutual scattering of two particles, should be found in gas
approximation. It is $\Gamma$, that should be used for estimation of
the interaction role within the gas limit instead of the initial
interaction $U$. For this purpose diagram technique is used and
calculations in stair approximation are made (sum frequency is equal
to double particle energy in the band bottom, total momentum is
zero):  $$\Gamma = U + 2i\ \frac{U^2}{N_0}\ <GG>+...=
\frac{U}{1-2i(U/N_0)<GG>}\ ;$$$$
\Gamma\bigl|_{U\rightarrow\infty}\rightarrow\ \frac{iN_0}{2<GG>}\
,$$
$$<GG>=\sum_{\bf p}\int \frac{d \omega}{2\pi}G(\Omega+\omega,{\bf
p})G(-\omega,-{\bf p})\ ,$$$$ G(\omega,{\bf
p})=\frac{1}{\omega-\epsilon({\bf p})+i\delta} \ .$$

Here sum frequency is $\Omega=2\epsilon(0)$. The result of the
calculation is: \begin{eqnarray}
\label{14}\Gamma^{-1}=\frac{-2i}{N_0}<GG> =\frac{1}{N_0} \sum_{\bf
p}\frac{1}{\epsilon({\bf p})-\epsilon(0)}\ ;\\ \nonumber
\Gamma^{-1}\rightarrow \frac{0.505}{2t}\ .\ \ \ \ \ \
\end{eqnarray} The last value is given for simple cubic lattice. The
expression for energy (all the particles are in the condensate
$A^+(0)=A(0)\rightarrow\sqrt{N}$) is: $$E\rightarrow
\epsilon(0)<A^+(0)A(0)> + \frac{\Gamma}{N_0}<A^+(0)A^+(0)A(0)A(0)>
$$$$=  -t\nu_0 N +\frac{\Gamma}{N_0}\ N^2=\ -t\nu_0 N\biggl(1
-\frac{2}{0.505\nu_0}\ \frac{N}{N_0}\biggr)\ .$$

One can see, that contribution of interaction for cubic lattice
($\nu_0 = 6$) is one and a half times less than for used trial
function. It should be emphasized, that it is the consequence of
binary collision approximation.

\subsection{Accuracy evaluation}
Corrections to energy can be estimated using the functions resulting
from Hamiltonian action on the function $\Phi_N$. Thus:
$$S^-(n)\Phi_N\equiv S^-(n)S^+\Phi_{N-1} =\Bigl[S^+S^-(n)-2S_z(n)
\Bigr] \Phi_{N-1}\ .$$ First, the value $S_z(n)\Phi_N$ is found. It
is easy to see, that: $$S_z(n)\Phi_N = \Phi_N(n) +
S^+\Bigl\{S_z(n)\Phi_{N-1} \Bigr\}\ .$$ From this recurrent relation
follows: \begin{eqnarray} \label{15}S_z(n)\Phi_N = N\Phi_N(n)
-\frac{1}{2} \Phi_N\ .\end{eqnarray} It is verified directly or by
summation by $n$. Thus:  $$S^-(n)\Phi_N = \Phi_{N-1} -
2(N-1)\Phi_{N-1}(n) + S^+ \Bigl\{ S^-(n)\Phi_{N-1} \Bigr\}\ .$$ From
this recurrent relation follows:  \begin{eqnarray} \label{16}
S^-(n)\Phi_N = N \Phi_{N-1} - N(N-1) \Phi_{N-1}(n)\ .\end{eqnarray}
It is verified by a direct substitution as well as at $N=1,\ N=2$.

The result is: \begin{eqnarray} \label{17} H \Phi_N =
-t\biggl\{N\nu_0 \Phi_N - N(N-1) \sum_{<nn'>}\Phi_N(n,n')\biggr\}
.\end{eqnarray} The first term arises from the particles at the band
bottom, the second - from interaction of these particles and
orthogonal to $\Phi_N$ states ($\Phi_{N\perp}$). The last should be
determined for corrections to the energy of an initial state to be
found. Noteworthily, that the same energy value (\ref{10}) is
obtained.

We may write:  \begin{eqnarray} \label{18} H\widetilde{\Phi}_N =
E\widetilde{\Phi}_N + w\widetilde{\Phi}_{N\perp}\ .\end{eqnarray}
Here $\widetilde{\Phi}_{N\perp}$ is a normalized function, $w$ is a
transition matrix element between the states $\widetilde{\Phi}_N$
and $\widetilde{\Phi}_{N\perp}$. This element should be found for
the corresponding two-level  problem to be considered.

Thus, according to (\ref{17}):  $$H\widetilde{\Phi}_N = -t
N\nu_0\biggl\{1-\frac{N-1}{N_0-1} \biggr\}\widetilde{\Phi}_N +$$$$+
t D_N N(N-1)\biggl\{ \sum_{<nn'>}
\Phi_N(n,n')-\frac{\nu_0}{N_0-1}\Phi_N\biggr\}\ .$$ The second term
is a sought quantity:  \begin{eqnarray} \label{19}
w\widetilde{\Phi}_{N\perp}=t D_N N(N-1)\Phi'_N \ ;\\
\nonumber \Phi'_N \equiv\biggl\{ \sum_{<nn'>} \Phi_N(n,n')
-\frac{\nu_0}{N_0-1}\Phi_N\biggr\}\ .\end{eqnarray}

First let us find a norm of function $\Phi'_N$:
$$\Bigl(\Phi'_N,\Phi'_N \Bigr) = \sum_{<m' n'>}\sum_{<m
n>}\Bigl(\Phi_N(m',n'),\Phi_N(m,n) \Bigr)-$$$$
-\frac{\nu_0^2}{(N_0-1)^2}\Bigl(\Phi_N,\Phi_N \Bigr)\ .$$ Here the
orthogonality of functions $\Phi_N$ and $\Phi'_N$ is used.

Various cases should be taken into account in the calculations: all
the numbers are different ($W_N^{(0)}$), two numbers are equal
($W_N^{(1)}$), two pairs of the coinciding numbers ($W_N^{(2)}$).
The result is the following: $$\sum_{<m' n'>}\sum_{<m
n>}\Bigl(\Phi_N(m',n'),\Phi_N(m,n) \Bigr) = $$$$=W_N^{(2)}\
2N_0\nu_0\ +\ W_N^{(1)}\ 4N_0\nu_0(\nu_0-1) +$$$$+ W_N^{(0)}\
\Bigl\{N_0^2\nu_0^2  - 4N_0\nu_0(\nu_0-1) - 2N_0\nu_0\Bigr\}\ .$$

Calculation of coefficients can be illustrated by example of
$W_N^{(0)}$. We have: $$W_N^{(0)}= \Bigl( \Phi_{N}(m'n'), \Phi_{N}(n
m ) \Bigr)\Bigl|_{\neq} =$$$$= \frac{1}{N_0-3}\sum_{l\neq m n}\Bigl(
\Phi_{N}(ln'), \Phi_{N}(n m ) \Bigr)\  = $$$$=\
\frac{1}{N_0-3}\biggl\{\sum_{l}\Bigl( \Phi_{N}(ln'), \Phi_{N}(n m )
\Bigr)-$$$$-\Bigl( \Phi_{N}(mn'), \Phi_{N}(n m ) \Bigr) - \Bigl(
\Phi_{N}(nn'), \Phi_{N}(n m ) \Bigr)\biggr\}\rightarrow$$
$$\rightarrow\ \frac{1}{N_0-3}\biggl\{ \Bigl( \Phi_{N}(n'),
\Phi_{N}(n m ) \Bigr) -2 W_N^{(1)}\biggr\} =$$$$=
\frac{1}{N_0-3}\Biggl\{ \frac{(N-2)\Bigl( \Phi_{N}, \Phi_{N}
\Bigr)}{NN_0(N_0-1)(N_0-2)} -2 W_N^{(1)}\Biggr\}\ .$$

Similar operations are made in the other cases:
\begin{eqnarray}\nonumber W_N^{(0)} = \frac{1}{N_0-3}\Biggl\{
\frac{(N-2)\Bigl( \Phi_{N}, \Phi_{N} \Bigr)}{NN_0(N_0-1)(N_0-2)} -2
W_N^{(1)}\Biggr\}; \\ \label{20} W_N^{(1)}=
\frac{1}{N_0-2}\biggl\{\frac{\Bigl( \Phi_{N}, \Phi_{N}
\Bigr)}{NN_0(N_0-1)} -W_N^{(2)}\biggr\}; \\ \nonumber W_N^{(2)}
=\frac{\Bigl( \Phi_{N}, \Phi_{N} \Bigr)}{N(N-1)N_0(N_0-1)}\
.\end{eqnarray}

The following relations for the sought quantity and for $w$, defined
in (\ref{19}), result from combining of all above mentioned
relations: \begin{eqnarray} \label{21} \Bigl(\Phi'_N,\Phi'_N \Bigr)
\approx 2\nu_0\ \frac{(N_0-N)^2}{N^2N_0^3}\ \Bigl(\Phi_N,\Phi_N \Bigr)\ ;\\
\nonumber w\approx  t \sqrt{2\nu_0}\  \frac{N(N_0-N)}{N_0^{3/2}}\ \
\ \ \ \ \ \ (N,N_0\rightarrow\infty)\ .\end{eqnarray}

Now the correction to the system energy can be estimated. The value
$w$ tends to infinity within the macroscopic limit, therefore it is
a fortiori much higher than the difference of $\widetilde{\Phi}'_N$
state energy and the initial one. Consequently, the correction to
energy is approximately equal to  $- w$. And this value is
proportional to the square root of volume, i.e.  is much lower than
the energy of interest (proportional to volume). Therefore, one can
draw a conclusion, that initial test function is a good
approximation of the problem.

For reliability, the energy root-mean-square value (i.e. $<H^2>$) is
found and compared with the value $<H>^2=E^2$. There are all the
data for the following relations:
$$\Bigl(\widetilde{\Phi}_N,H^2\widetilde{\Phi}_N \Bigr) =
\Bigl(E\widetilde{\Phi}_N+w\widetilde{\Phi}_{
N\bot},E\widetilde{\Phi}_N+w\widetilde{\Phi}_{ N\bot} \Bigr)=
E^2+w^2  ;$$ \begin{eqnarray} \label{22}\frac{<(H-E)^2>}{E^2} =
 2\nu_0 t^2 \frac{N^2(N_0-N)^2}{E^2N_0^3}\biggl|_{N,N_0\rightarrow\infty} \rightarrow 0
.\end{eqnarray} Here a finite concentration is meant (the ratio
$N/N_0$ is constant).

In (\ref{22}) the system ground state is shown to be well described
by the trial function used.  It is true for an $X-Y$ model in
general.

\section{Conclusion}

Within the framework of the model used, the conventional Bose gas
theory for the particles with strong interaction was shown to be
inconsistent. If most particles are in Bose condensate, description
of their interaction based on binary collisions (using binary
scattering amplitude) does not suit, as there is no quasi-classics
for the particles with an infinite wave length.  It turns out, as if
every particle interacts with all the particles at once. This fact
leads to the increase of the interaction energy, as shown in the
model used here.

It should be emphasized, that in present work the conclusion about
the accuracy of the approach is made after writing of a trial
function. In \cite{3} first the problem is simplified by a
transition to a pseudo-potential, then solved by the perturbation
theory. However, there are no successful attempts to write a
multi-particle function for initial interaction (hard spheres).

Note, that the results obtained suit for two-dimensional case
(three-dimensional spin).

So far it is not clear, how the spectrum of elementary excitations
can be got. Perhaps, a diagram Belyaev - type technique (see, for
example, \cite{5}) or its modified form (?) should be used.

Acknowledgement. The author gratefully acknowledges the discussion
of  A.V. Chaplik, M.V. Entin  and V.M. Kovalev and also the
financial support of RFBR (Grant 11-02-00060) and Russian Academy of
Sciences (Programs).

\end{document}